\RequirePackage[colorlinks = true, urlcolor   = blue, citecolor  = blue]{hyperref}
\documentclass[draft,grl]{agujournal2019}
\usepackage{url} %this package should fix any errors with URLs in refs.
\usepackage{lineno}
\usepackage{natbib}
\usepackage[inline]{trackchanges} %for better track changes. finalnew option will compile document with changes incorporated.
\usepackage{soul}
\draftfalse

\journalname{Geophysical Research Letters}
\usepackage{amsmath}
\usepackage[inkscapeformat=png]{svg}
\usepackage{calrsfs}
\newcommand\Rey{\mbox{\textit{Re}}}  % Reynolds number
\newcommand\Web{\mbox{\textit{We}}}  % Weber number
\newcommand\Sc{\mbox{\textit{Sc}}}  % Schmidt number
\newcommand{\diff}{\mathrm{d}}

\usepackage{color}
\definecolor{ForestGreen}{RGB}{0,128,0}

\usepackage{graphicx}
\usepackage{tikz}
\usetikzlibrary{positioning} % Optional, but useful for node positioning
\usepackage{xr}
\begin{document}

\title{Air entrainment and gas transfer \\ in wave breaking events}

%%%%%%%%%%%%%%%%%%%%%%%%%%%%%%%%%%%%%%%%%%%%%%%
%
%  AUTHORS AND AFFILIATIONS
%
%%%%%%%%%%%%%%%%%%%%%%%%%%%%%%%%%%%%%%%%%%%%%%%

% Authors are individuals who have significantly contributed to the
% research and preparation of the article. Group authors are allowed, if
% each author in the group is separately identified in an appendix.)

% List authors by first name or initial followed by last name and
% separated by commas. Use \affil{} to number affiliations, and
% \thanks{} for author notes.
% Additional author notes should be indicated with \thanks{} (for
% example, for current addresses).

% Example: \authors{A. B. Author\affil{1}\thanks{Current address, Antartica}, B. C. Author\affil{2,3}, and D. E.
% Author\affil{3,4}\thanks{Also funded by Monsanto.}}

\authors{S. Di Giorgio\affil{1,2},
 S. Pirozzoli\affil{3},
 A. Iafrati\affil{1}}

\affiliation{1}{Institute of Marine Engineering - CNR - Rome, Italy}
\affiliation{2}{Institute of Fluid Mechanics and Heat Transfer - TU-Wien, Austria}
\affiliation{3}{Department of Mechanics and Aerospace Engineering - Sapienza Univ. Rome,
Italy}

% Corresponding author mailing address and e-mail address:

% (include name and email addresses of the corresponding author.  More
% than one corresponding author is allowed in this LaTeX file and for
% publication; but only one corresponding author is allowed in our
% editorial system.)

% Example: \correspondingauthor{First and Last Name}{email@address.edu}

\correspondingauthor{Simone Di Giorgio}{simone.digiorgio@cnr.it}

%%%%%%%%%%%%%%%%%%%%%%%%%%%%%%%%%%%%%%%%%%%%%%%
% KEY POINTS
%%%%%%%%%%%%%%%%%%%%%%%%%%%%%%%%%%%%%%%%%%%%%%%
%  List up to three key points (at least one is required)
%  Key Points summarize the main points and conclusions of the article
%  Each must be 140 characters or fewer with no special characters or punctuation and must be complete sentences

% Example:
% \begin{keypoints}
% \item	List up to three key points (at least one is required)
% \item	Key Points summarize the main points and conclusions of the article
% \item	Each must be 140 characters or fewer with no special characters or punctuation and must be complete sentences
% \end{keypoints}

\begin{keypoints}

\item High-fidelity simulations show that wave breaking boosts gas transfer across the air-water interface due to air entrainment and turbulence.

\item Gas transfer velocity scales with one-fourth power of the energy dissipation rate, consistent with theoretical predictions.

\item Precise estimates of gas transfer velocity are achieved enabling a more accurate modeling of oceanic CO\(_2\) absorption.

\end{keypoints}

\justifying

%%%%%%%%%%%%%%%%%%%%%%%%%%%%%%%%%%%%%%%%%%%%%%%
%
%  ABSTRACT and PLAIN LANGUAGE SUMMARY
%
% A good Abstract will begin with a short description of the problem
% being addressed, briefly describe the new data or analyses, then
% briefly states the main conclusion(s) and how they are supported and
% uncertainties.

% The Plain Language Summary should be written for a broad audience,
% including journalists and the science-interested public, that will not have 
% a background in your field.
%
% A Plain Language Summary is required in GRL, JGR: Planets, JGR: Biogeosciences,
% JGR: Oceans, G-Cubed, Reviews of Geophysics, and JAMES.
% see http://sharingscience.agu.org/creating-plain-language-summary/)
%
%%%%%%%%%%%%%%%%%%%%%%%%%%%%%%%%%%%%%%%%%%%%%%%

%% \begin{abstract} starts the second page

\begin{abstract}
We investigate gas transfer processes occurring at the air-water interface of progressive 
water waves using high-fidelity numerical simulations. Waves with varying initial 
steepness, including regular wave patterns, mild spilling and intense plunging breakers 
are examined. A two-phase solver is employed to model exchange processes enabling 
precise estimation of the air-water interface area and gas transfer velocity, achieving 
an accuracy unattainable in experiments. We show that the volume of gas transferred 
across the air-water interface increases significantly with the amount of air entrained 
due to wave breaking, peak values in the transfer velocity being concurrent with peaks 
in energy dissipation rate and air entrainment. The gas transfer velocity is observed 
to scale approximately as the one-fourth power of the energy dissipation rate, consistent 
with previous theoretical predictions. The present findings can help reduce the 
substantial uncertainty associated with the parametrization of fundamental natural 
processes, such as CO$_2$ absorption by the oceans.

\end{abstract}

\section*{Plain Language Summary}
When ocean waves break, the entrainment of air bubbles and turbulence enhances the transfer of gases, such as oxygen and carbon dioxide, across the ocean-atmosphere interface. This process governs Earth's climate and controls marine ecosystems. However, accurately estimating the amount of gas transferred during wave-breaking events remains challenging.  
We employ high-fidelity numerical simulations to investigate gas transfer during breaking of waves with varying intensities. These simulations provide a level of detail and accuracy that is difficult to achieve through laboratory experiments.  
We find that the intensity of wave breaking is directly correlated with the amount of gas transferred across the interface and the gas transfer velocity. The results show that gas exchange is closely tied to air entrainment and the induced turbulence, which increases the rate of energy dissipation.  
We observe that the gas transfer velocity scales with the energy dissipation rate raised to a power of one-fourth, aligning with existing theoretical predictions.  
This research contributes to reducing uncertainties in the estimation of gas exchange under natural conditions, such as the capacity of the oceans to absorb carbon dioxide. By improving the accuracy of climate models, these findings enable more reliable predictions of the ocean's response to global environmental changes.

\section{Introduction}

Gas exchange processes at the air-sea interface play a crucial
role in regulating the climate and sustaining both human and marine life.
A significant portion of anthropogenic carbon dioxide is absorbed
by the ocean~\citep{watson2020,friedlingstein2020global}, which, in turn,
releases nearly half of the oxygen we breathe through the photosynthesis of
marine flora in the sunlit upper ocean layer.
For low-solubility gases such as oxygen, mass transfer is governed by molecular
diffusion and turbulence within a very thin layer on the water
side~\citep{jahne1998arfm,Jirka2010}.
Although the original motivation for the study stems from the exchange
processes at the ocean surface, the gas transfer across a
gas-liquid interface is of great interest in other contexts such as
chemical, food and pharmaceutical industries where bubble columns are often
used in chemical reactors~\citep{deising2018}.

Notwithstanding its importance, gas exchange processes
remain scarcely understood~\citep{deike2022mass}. Most studies have focused
on correlating gas transfer velocity with wind speed~\citep[e.g.][]{wanninkhof2009}, 
yet the underlying mechanisms driving gas
exchange processes are not fully elucidated.
One of the reasons is that laboratory measurements are extremely challenging as
concentration fluctuations should be measured at depths of at most
hundreds of micrometers to have direct relevance to air–water gas
transfer~\citep{asher2009}, hence most investigations
pertain to unbroken air-sea interfaces~\citep[e.g.][]{chu2003, herlina2004}.
Despite the clear evidence that bubbles generated from wave breaking,
with associated turbulence and energy dissipation,
enhance significantly air-sea exchanges especially for
poorly soluble gases, the parameterization of their effect
is grossly inaccurate~\citep{garbe2014}.

The gas transfer velocity is often expressed in terms of the
near-surface turbulent dissipation rate~\citep{kitaigorodskii1984jpo,shuiqing2016}).
Since air entrainment and bubble fragmentation are known to
significantly promote energy dissipation~\citep{iafrati2011},
a similar effect on gas transfer is expected.
The effect of air bubbles on mass diffusion involves two distinct cases:
smaller bubbles that lack sufficient buoyancy to rise and
gradually dissolve into the water, and larger bubbles that dissolve
partially while ascending and eventually burst at the free surface
~\citep{stanley2009}.
Measurements of the gas transfer velocity under breaking waves,
induced via modulational instability with and without overlying wind,
were conducted by~\citet{li2021jpo}.
Those authors proposed that a Reynolds number, based on the breaker
height and the mean orbital velocity of the breaking wave,
is a relevant parameter.
While it is anticipated that these parameters influence the
bubble injection rate~\citep{iafrati2011}, no direct experimental
evidence was provided. Additionally, in those experiments,
the gas concentration was measured only before and after breaking,
leaving the temporal evolution of the local concentration unobserved.

The significant progress of multiphase flow numerical solvers in the last
fifteen years along with recent introduction of numerical methods to model gas
transfer processes across gas-liquid
interfaces~\citep{deising2018,farsoiya2021bubble},
have made it possible to investigate numerically the bubble-mediated
contribution to the gas transfer.
An attempt in this direction was made by \citet{mirjalili2022ijhmt},
who studied the dissolved gas concentration in a two-dimensional breaking wave,
with a somewhat unrealistic air-water density ratio of $0.01$.
However, no quantitative data about gas transfer at the interface was provided.

\section{Methods}
In order to investigate gas transfer processes taking place during the breaking
of free-surface waves and to identify the significance of air entrainment,
herein we numerically simulate the time evolution of a third-order Stokes'
wave~\citep{deike2015capillary, digiorgio2022jfm, mirjalili2022ijhmt},
for various initial steepness, yielding a regular wave pattern,
mild spilling breaking, and intense plunging breaking with substantial air
entrainment.
The flow is assumed to be three-dimensional and periodic
along the streamwise ($x$) and spanwise ($z$) directions, $y$ being the
vertical axis.
We solve the Navier-Stokes for an incompressible fluid with variable fluid
properties
across the air-water interface, which we model after a geometric Volume-of-Fluid
method~\citep{weymouth2010conservative}. A detailed description of the
solver and its validation is provided 
in~\citet{digiorgio2022jfm,digiorgio2024jcp}.

The baseline multiphase solver is here augmented with a model
to account for gas diffusion, whereby
the time evolution of the gas concentration $c_{w/a}$ in water ($w$)
and air ($a$) is determined by solving~\citep{standart1964mass, 
haroun2010volume}
\begin{linenomath*}
\begin{equation}
\label{eq:concentration}
\frac{ \partial c_{w/a} }{\partial t }
+ \nabla \cdot \left( \mathbf{u} c_{w/a} \right) =
- \nabla \cdot \left(\mathbf{J}_{w/a} \right),
\end{equation}
\end{linenomath*}
where $\mathbf{u}$ is the local fluid velocity,
and $\mathbf{J_{w/a}}$ is the gas flux vector.
The standard assumption of continuous chemical potentials at the
air-water interface leads to Henry's law~\citep{bothe2013volume},
expressed as
\begin{linenomath*}
\begin{equation}
c_{w} = \alpha c_{a},
\end{equation}
\end{linenomath*}
where $\alpha$ is the solubility constant~\citep{sander2023compilation},
assumed constant in this study.
We use the water fraction ($\chi$) to evaluate the local gas
concentration as a weighted average of the values in air and water.
Similarly, the local diffusivity coefficient is evaluated as a
harmonic mean~\citep{deising2016unified}. These formulations lead to
the nondimensional form of the concentration equation,
\begin{linenomath*}
\begin{equation}
\label{eq:concentrazione}
\frac{ \partial c }{ \partial t }
+ \nabla \cdot \left( \mathbf{u} c \right) = \frac{1}{\Rey \Sc}
 \nabla \cdot \left( D \nabla c - D
\left( \frac{ c \left(\alpha-1 \right) } { \alpha \chi + \left( 
1-\chi \right)}    \right)   \nabla \chi \right) \;\;.
\end{equation}
\end{linenomath*}
The convective term in the above equation is discretized with an upwind-biased
TVD scheme~\citep{pirozzoli2019}. 
Convergence and validation tests of the numerical model adopted for 
the gas transfer have been conducted for diffusion from static and
rising bubbles, with results in agreement with
the existing literature~\citep{digiorgio2025fedsm}.

A third-order wave is considered as initial condition~\citep{iafrati2009numerical, deike2015capillary,digiorgio2022jfm}, whose profile is specified as
\begin{linenomath*}
\begin{equation}
\eta (x,z) = \frac{\epsilon}{2 \pi} \left( \cos ( k x')
+ \frac{\epsilon}{2} \cos ( 2 k x') + \frac{3 \epsilon^2}{8} 
\cos ( 3 k x') \right) \;\;,
\end{equation}
\end{linenomath*}
where the wavelength $\lambda$ is hereafter assumed to be the reference length,
$k = 2 \pi/\lambda$ is the fundamental wavenumber, $\epsilon = a k$ is the
initial wave steepness, and $x' \approx x$, unless
small random perturbation~\citep{digiorgio2022jfm}.
From wave theory, assuming $U_R = (g \lambda)^{1/2}$ as reference velocity
and $T_R = (\lambda/g)^{1/2}$ as reference time,
the nondimensional period of the fundamental wave component is
$T_p = (2 \pi)^{1/2}$.
No-slip boundary conditions are enforced at the top and bottom boundaries.
The initial velocity in the water domain ($y < \eta(x,z)$) is
determined from second-order potential flow theory, whereas the
air side is assumed at rest.
The gas is initially assumed to be at the saturation point in the air
domain and absent in the water domain, as shown in the top left
panel of Figure~\ref{fig:conc_cont}.
Although this somewhat unphysical setup yields 
some spurious initial transient, we decided to adopt it as 
it minimizes the number of free parameters.

The energy content in water ($E_w$) is evaluated as the sum of the kinetic
and potential contributions, as follows
\begin{linenomath*}
\begin{equation}
E_w(t) = \int_{V_w} \rho \left( \frac{|{\mathbf u}|^2}{2} + g y\right)
\diff V - E_{p0} \;\;,
\end{equation}
\end{linenomath*}
where $E_{p0}$ is the potential energy of the fluid at rest.

The numerical simulations are carried out for Weber number
$\Web = (\rho_w U_R^2 \lambda)/\sigma = 12,000$, with $\sigma$ the surface
tension coefficient, which corresponds to waves with about $30$~cm
fundamental wavelength. At such scale, the Reynolds number
$\Rey = (\rho_w U_R \lambda)/\mu_w$ would be about $500,000$,
too high for all the scales to be fully resolved. Hence,
numerical simulations are carried out at reduced Reynolds numbers $\Rey =
10,000$ and $\Rey = 40,000$, which correspond to fundamental wavelengths of
$2.17$~cm and $5.46$~cm, respectively.
Although rather short, these waves are not too far from
those observed at the lowest wind speed in \citet{zappa2001}.

Three values of the initial steepness are considered, $\epsilon = 0.25, 0.37$
and $0.50$, which lead, respectively, to a regular wave pattern, mild spilling
breaking with small air
entrainment, and intensive wave breaking with large air entrainment.
The computational domain is one fundamental wavelength long, two wavelengths
tall and half wavelength wide.
It is discretized by using $N_x = 1152$, $N_z = 576$, $N_y = 768$ collocation points, 
uniformly distributed in the $x$ and $z$ directions, and in the $y$ direction 
between $y = -\lambda/4$ and $y = +\lambda/4$. The grid size is the same in all directions
in the uniform grid region, whereas the cell size increases towards the
bottom and the top boundaries according to a hyperbolic tangent function~\citep[][sec.~2.2.3b]{orlandi2012fluid}.
Assuming $\lambda = 30$~cm, the resulting grid size in the
well-resolved zone is $0.26$~mm.

As for the dissolving gas, we assume the Schmidt number in water to be 
$Sc_w = \mu_w/(\rho_w D_w) = 4$, the diffusivity ratio to be
$D_a/D_w = 100$, and the solubility constant to be $\alpha = 0.33$.
The use of low Schmidt number is motivated by the need of keeping
the grid points within acceptable bounds~\citep{nagaosa2014}.
However, extrapolation of the gas transfer velocity to higher Schmidt numbers
is possible by using the Levich law~\citep{levich1962physicochemical}.

\begin{figure}[hbt]
\includegraphics[width=14.5cm]{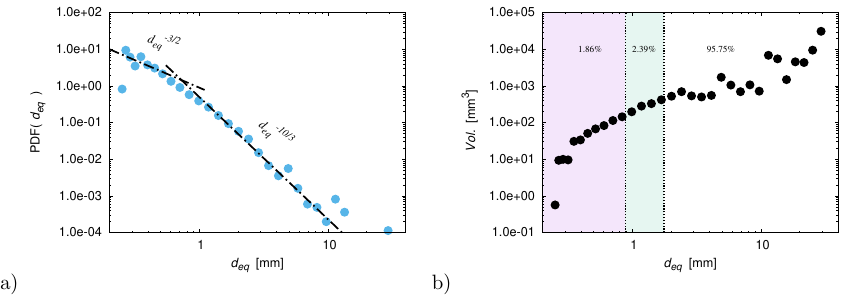}
\caption{\label{fig:bub_resol}
(a) Bubbles size distribution for the simulation at $\Rey = 40,000$ and 
steepness $\epsilon = 0.50$. (b) Distribution of bubble volume as a function 
of equivalent radius. 
The volume percentage of bubbles is categorized into three ranges:
$d_0 < 12$; $12 <d_0/\Delta x < 25$;  $d_0/\Delta x > 25$.
The bubble statistics are computed using collective data over the time interval 
$t/T_p \approx 0.75$ to $t/T_p \approx 1.5$, which corresponds to the most intense 
breakup phase of the simulation.
 }
\end{figure}

The adequacy of the grid resolution for the description of the
bubble dynamics was already verified in a previous study~\citep{digiorgio2022jfm}. 
In Figure~\ref{fig:bub_resol}a, the bubble size distribution is shown
for the case with $\Rey=40,0000$ and $\epsilon = 0.50$, which is the case 
yielding to the smallest bubbles, and it exhibits the trends with the
two different power laws, as found by \citet{deane2002}.
Concerning with the gas transfer, from a careful validation study performed
in \citet{digiorgio2025fedsm}, it has been found that, for bubbles with a 
diameter smaller than 25 grid cells, the gas transfer velocity is overestimated 
by approximately 8\%.
However, based on the bubble size distribution shown in Figure~\ref{fig:bub_resol}a, 
the total volume of the bubbles with a diameter smaller than 25 grid cells 
accounts for less than 4\% of the total air volume entrained by wave breaking
(Figure~\ref{fig:bub_resol}b) and, therefore, the effect on the total gas transfer
is negligible.

Starting from the local concentration, the total amount of gas in 
air ($q_a$) and water ($q_w$) can be determined from integration,
\begin{linenomath*}
\begin{equation}
\label{eq:masw}
 q_{a,w} = \int_{V_{a,w}} c(x,y,z) \, \diff V \;\;,
\end{equation}
\end{linenomath*}
where $V_{a,w}$ is the volume occupied by the two phases.
Mean concentrations of gas in air and water can then be defined as
$\bar{c}_{a,w} = q_{a,w}/V_{a,w}$.

Due to the finite grid resolution 
that can be achieved in numerical simulations 
and to the lack of a suitable dissolution
model~\cite[e.g.][]{farsoiya2023}, the asymmetric contribution to gas 
transfer associated with the finest bubbles~\citep{stanley2022, zhang2012asy},
is not accurately accounted for.
Although this could be an important shortcoming,
as metioned in~\citet{deike2022mass,emerson2019air}, this contribution to gas transfer 
is more relevant for low-solubility gases like N2 and O2, 
with solubility values of $\alpha = 0.012$ and $0.025$, respectively. 
It is not expected to substantially affect the conclusions of our study, 
in which the solubility is an order of magnitude higher.

\section{Results}

\begin{figure}
\centering
\includegraphics[width=12.5cm]{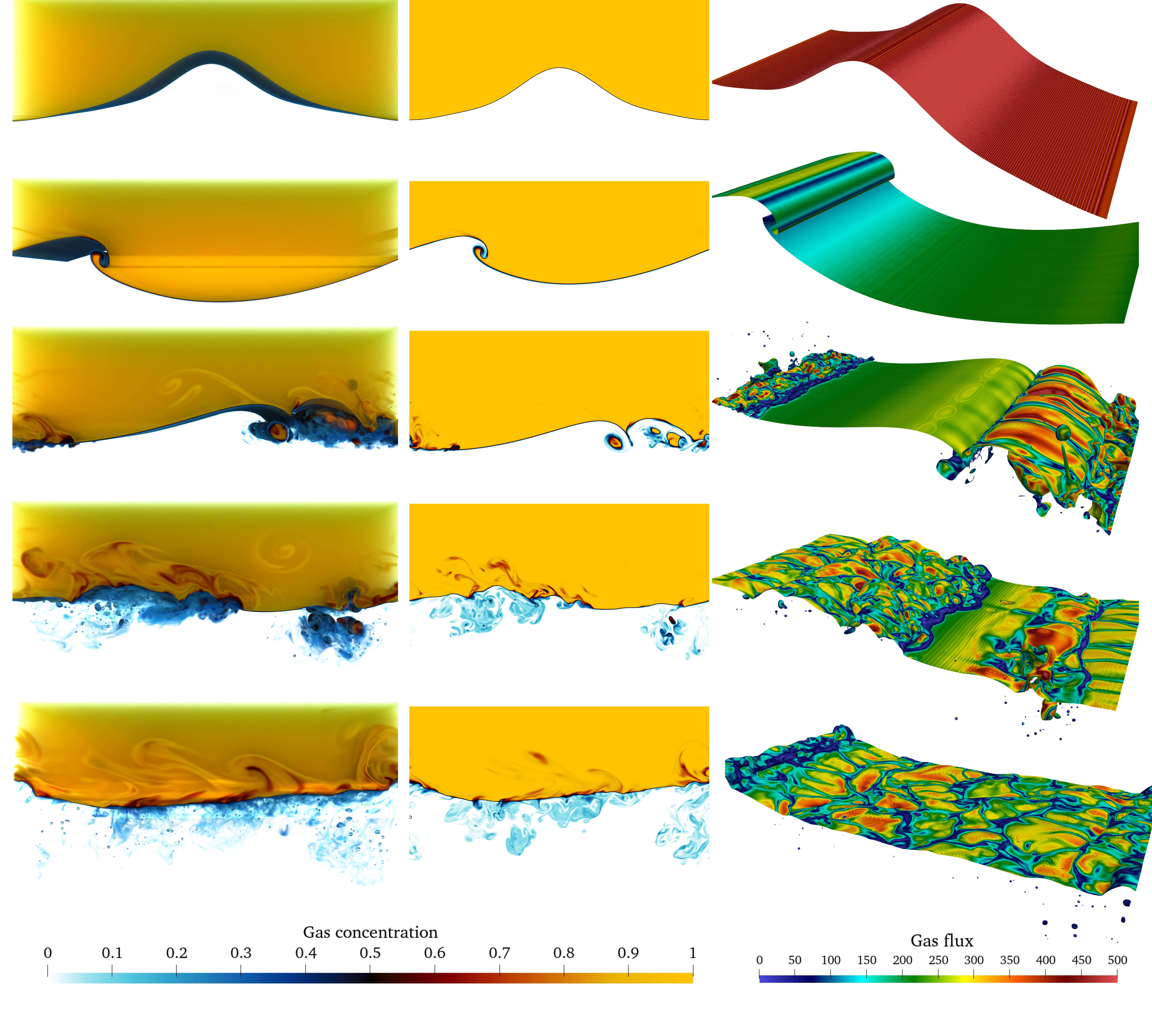}
\caption{\label{fig:conc_cont}
Sequences of 3D rendering of the concentration distribution (left)
gas concentration contours on a slice (center) and gas flux at the air-water
interface (right).
The data refer to the flow case with $\epsilon = 0.50$, $\Rey=40,000$.
From top to bottom $t/T_p = 0, 0.50, 1.00, 1.50, 2.00$.}
\end{figure}

In Figure~\ref{fig:conc_cont}, we present the three-dimensional rendering of the 
gas concentration together with the free surface, the contours of the gas
concentration on a vertical slice, and the gas flux normal to the air-water interface.
Results are presented at various
stages of wave evolution for the case $\epsilon = 0.50$, $\Rey = 40,000$,
which results in maximum gas transfer.
The abrupt jump in gas concentration imposed as the initial condition
causes a high gas flux through the interface at the start of the
simulation, as shown in the top-right panel. This spurious transient
is completed before the onset of breaking ($t/T_p \approx 0.5$).
The sequence clearly illustrates that, during the early stages of the
breaking process, gas transfer predominantly occurs across the air-water
interface, with gas becoming trapped in air bubbles entrained in the water.
At later stages, the entrapped gas diffuses into the water domain,
accompanied by a corresponding decrease in concentration in the air domain.
The three-dimensional renderings highlight how air entrainment and bubble
fragmentation enhance gas transfer and diffusion in the water domain. 

\begin{figure}[hbt]
%a) \hspace{-1.cm}\includegraphics[width=7.5cm]{fig2a}
%b) \hspace{-1.cm} \includegraphics[width=7.5cm]{fig2b}
%\\
%c) \hspace{-1.cm}\includegraphics[width=7.5cm]{fig2c}
%d) \hspace{-1.cm}\includegraphics[width=7.5cm]{fig2d}
\includegraphics[width=14.5cm]{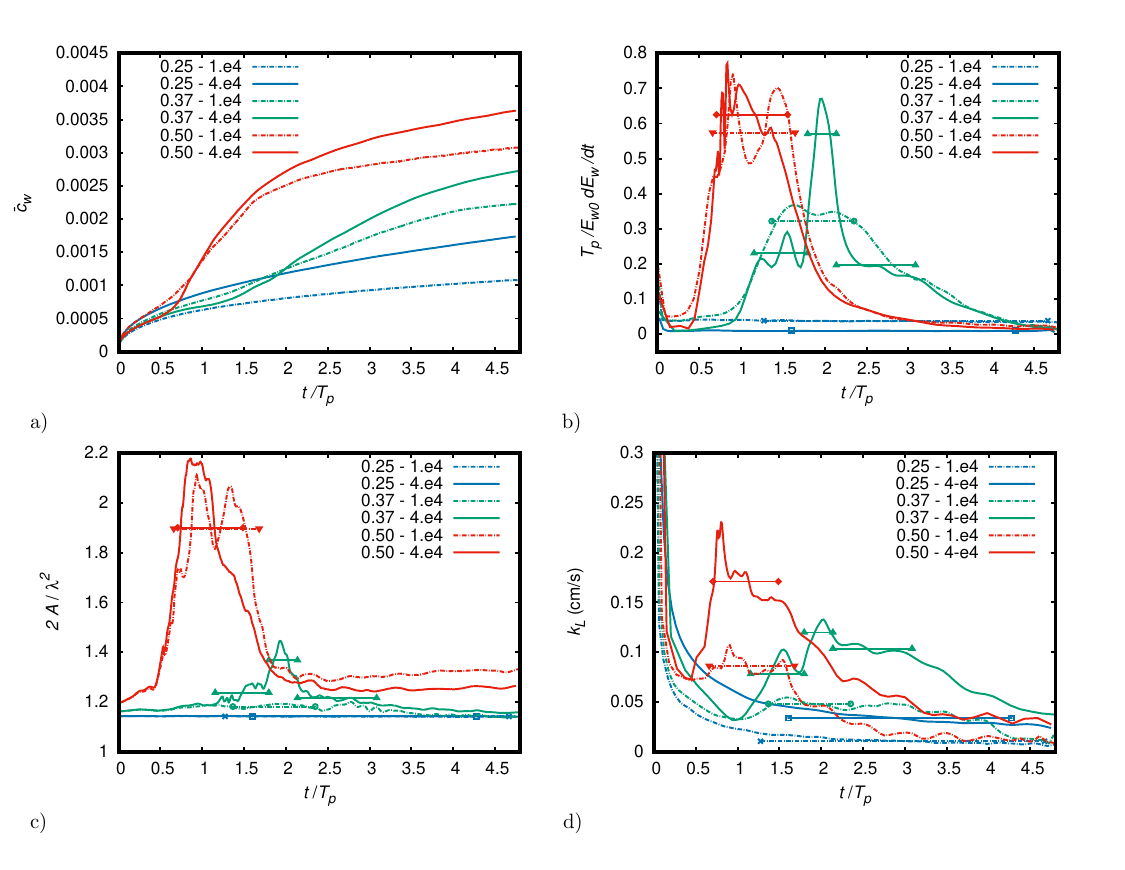}
\caption{\label{fig:conc_dirate}
Time history of mean gas concentration in water (a), of the
non-dimensional energy dissipation rate in water (b), 
of the overall air-water interface area,
normalized by the $x-z$ plane projected area ($\lambda^2/2$) (c) and 
of the gas transfer velocity, as defined in equation
\eqref{eq:kl} (d),
for different values of the initial steepness and Reynolds number.
The horizontal bars in panel (b,c,d) indicate the average
values during the various phases of the breaking and the time
intervals over which the averages are taken.}
\end{figure}

To quantify the gas transfer process, Figure~\ref{fig:conc_dirate} shows
the mean gas concentration in water, the energy dissipation rate, 
the overall air-water interface area (normalized by $x-z$ plane 
projected area) and the gas transfer velocity \eqref{eq:kl}
as functions of time.
The gas concentration exhibits an initial transient lasting about half
a wave period, which is similar for all cases. This phase is
characterized by intense gas transfer across the air-water interface,
driven by the artificial start-up, followed by a milder growth phase
with a typical time scale associated with the wave orbital velocity.
For the flow cases with $\epsilon=0.25$, featuring a regular wave
pattern, the growth rate of the mean gas concentration in water
progressively decreases with time.
In contrast, for the flow cases with $\epsilon=0.50$, a sudden
increase in the mean gas concentration is observed starting at
about half a wave period. This increase coincides with the sharp rise
in the energy dissipation rate, as shown in Figure~\ref{fig:conc_dirate}b,
marking the onset of the plunging breaking event. Intense gas transfer
persists up to $t \approx 2 T_p$, after which the mass transfer rate
returns to values similar to the non-breaking cases.
In milder spilling breaking cases, breaking begins shortly before
$t = T_p$, as evidenced by the increase in both gas transfer and
energy dissipation rates. In these cases, the breaking process lasts
longer, resulting in a gas transfer rate significantly lower than that
in the plunging breaking cases. The rate approaches the non-breaking
value around $t \simeq 4 T_p$.

The results reported in Figure~\ref{fig:conc_cont} clearly convey that air
entrainment plays an important role in the gas exchange process, as
also observed by several previous authors~\citep{memery1985, asher1996},
and as pointed out in recent reviews of the subject~\citep{wanninkhof2009,
deike2022mass}.
Our high-fidelity numerical model enables to make these qualitative statement
into quantitative predictions by quantifying the actual area
of the air-water interface ($A$) during wave breaking, which
also accounts for the surface area of bubbles, sprays and droplets.
This is evaluated by summing the areas of planar 
segments reconstructed using the PLIC-VOF method, as proposed 
by~\citet{aniszewski2021}.
The time histories of the overall air/water interface area
are shown in Figure~\ref{fig:conc_dirate}c. The data
display increase of the interface area also for the spilling
breaking case with $\epsilon = 0.37$ and $\Rey = 40,000$.
However, much more significant increase is observed for the plunging
breaking cases with $\epsilon = 0.50$, for which the
interface area increases by up to a factor two from the initial value.
The data in Figure~\ref{fig:conc_dirate}c and Figure~\ref{fig:conc_dirate}b display the strong correlation between the air entrainment and the
increase in the energy dissipation rate, as already noted by \citet{iafrati2011}.

Various theories have been proposed to explain and parameterize
gas exchange processes in air-water systems~\cite[e.g.][]{jahne1998arfm}.
By appealing to the "surface renewal" theory, \citet{lamont1970} proposed
that the gas transfer velocity is related to the turbulence energy
dissipation rate near the air-water interface, raised to the 1/4 power.
\citet{kitaigorodskii1984jpo} reached a similar conclusion by modeling the
influence of turbulence patches enhanced by wave breaking~\citep[see, e.g.][]{zappa2007grl}). 
Directly measuring the turbulence induced by
wave breaking is challenging~\citep{shuiqing2016, asher2009}. To
address this limitation, \citet{shuiqing2016} proposed using the dissipation
rate of wave energy as a proxy for the turbulent dissipation rate. Since wave
breaking both dissipates energy and generates turbulence near the air-sea
interface, they assumed that the turbulent dissipation rate
is proportional to the wave energy dissipation rate. Following a
similar line of thought, in our study we assume that the average turbulent dissipation
rate within the computational volume can be related—through an appropriate
proportionality constant—to the wave energy dissipation rate.

The results shown in Figure~\ref{fig:conc_dirate} indeed
corroborate these statements, confirming that the gas transfer rate is strongly
correlated with the energy dissipation rate.
Quantitative evaluation of the gas transfer rate can be made in terms of the
gas flux per unit surface, namely $J = 1/A \; \diff q_w/\diff t$,
with $q_w$ defined in equation~\eqref{eq:masw}.
Following \citet{wanninkhof2009}, we define the gas transfer velocity as

\begin{linenomath*}
\begin{equation}
\label{eq:kl}
k_L = \frac{1}{A} \frac{\diff q_w/\diff t} {\left( \bar{c}_w - 
\alpha \bar{c}_a \right)} \;\;,
\end{equation}
\end{linenomath*}
where the mean gas concentrations in the two fluids are used.

The time histories of the gas transfer velocity are shown
in Figure~\ref{fig:conc_dirate}d.
All plots display unnaturally large values in the initial stages owing to the
start-up transient.
Afterwards, the gas transfer velocity increases significantly with
the initial wave steepness and the breaking intensity,
which is consistent with what found for the average concentration in
Figure~\ref{fig:conc_dirate}a.
The data also indicate that for waves with same initial steepness the gas
transfer velocity increases when increasing the Reynolds number, hence with the
dimensional
wavelength, in agreement with the observations of~\citet{li2021jpo}.
Comparing the time histories with the results of Figure~\ref{fig:conc_dirate}b
and Figure~\ref{fig:conc_dirate}c, we further note that the gas transfer velocity
attains its peak at about the same time as the energy dissipation
rate and the air-water interface area.
The fact that $k_L$, which is evaluated by using the overall
air/water interface area,
attains its maximum value when $A$ is also maximum,
suggests that the increase in the gas flux is much stronger than due
to the sole increase of the air/water interface area.
In other words, not only does air entrainment widen the interface area
through which the gas exchange takes place, but it also
enhances the gas transfer velocity by magnifying the
velocity gradients occurring around the air/water interface as a consequence
of bubble fragmentation processes,
as highlighted from the concentration fields and the three-dimensional rendering in Figure~\ref{fig:conc_cont}.

It is worth remarking that the estimation of gas transfer velocity, as
defined in equation~\eqref{eq:kl}, relies on the overall air-water interface
area. While this quantity
is readily available in numerical simulations, laboratory or field
experiments typically estimate \(k_L\) using the waterplane area as a
surrogate,
i.e., the horizontal projection of the air-water interface assuming it to be flat.
As shown in Figure~\ref{fig:conc_dirate}c, this approximation can
be inadequate, particularly for energetic plunging breaking cases, where
the interface area can reach up to twice its initial value and more than
double the waterplane area, which for the present simulations is
$\lambda^2/2$.
Numerical simulations thus provide an unprecedented opportunity to
accurately evaluate the energy dissipation rate and gas transfer
velocity based on the overall interface area. This enables validation
of power-law formulas currently in use~\citep{shuiqing2016}. 
\begin{figure}
\centering
\includegraphics[width=12cm]{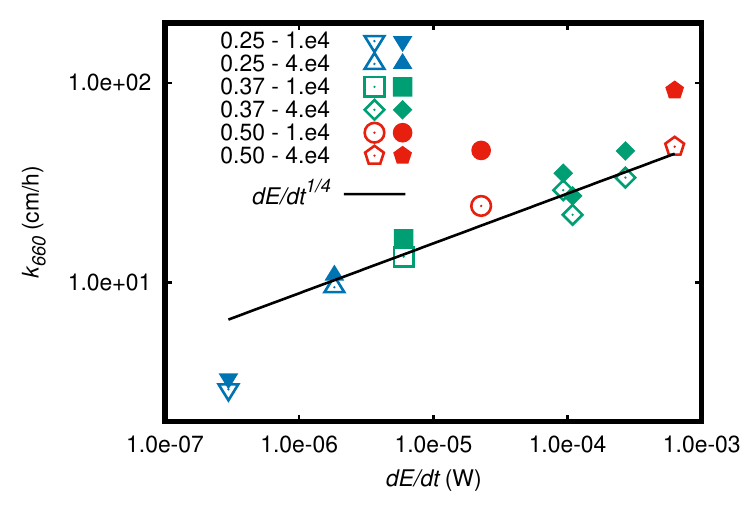}
\caption{\label{fig:kldiss}
Gas transfer velocity based on the overall air-water
interface (open symbols) and based on the waterplane area (solid symbols),
as a function of the respective dissipation rate.
Following standard practice in the field, the gas transfer velocity is
expressed in cm/h and extrapolated to \(\Sc = 660\) using the
\(Sc^{-1/2}\) scaling law \citep{levich1962physicochemical}.
The line represents the $1/4$ power of the energy dissipation rate.}
\end{figure}
The data for the various cases are presented in Figure~\ref{fig:kldiss}.
Despite noticeable dispersion in the data points, the gas transfer
velocity computed using the overall interface area (open symbols)
exhibits a clear increasing trend, consistent with the \(1/4\)
power-law model. Conversely, the velocity estimated using the
waterplane area (filled symbols) is slightly higher than that based
on the overall interface area for most cases. However, for the two cases with
plunging breaking, the values based on the waterplane area
are nearly double the true values.

Given the small scales simulated in this study, the gas transfer velocity  
values shown in Figure~\ref{fig:kldiss} appear relatively high compared to  
those reported in the literature based on laboratory experiments and  
wind-generated waves (e.g., \citep{wanninkhof2009}). Among the relevant studies,  
\citet{zappa2001, zappa2004} conducted experiments beginning at low wind speeds  
that produced microbreakers with minimal, if any, air entrainment. Two tracer  
gases, He and SF\textsubscript{6}, were used, and the influence of surfactants  
was also examined.
At the lowest wind speed tested, 4.6 m/s, the peak wave frequency was in the  
range of 4.8-5 Hz, corresponding to a wavelength of approximately 6.2-6.7 cm   
comparable to the 5.46 cm wavelength in our simulations at \(\Rey = 40000\).  
Under these experimental conditions, the gas transfer velocity, normalized to  
\(\Sc = 660\), was found to lie between 2.1 and 5.1 cm/h.
The data further revealed a linear relationship between the gas transfer  
velocity and the active breaking area \(A_B\), expressed as a fraction of the  
total surface area. The active breaking area was estimated using infrared  
imaging, which enabled identification of regions with enhanced heat transfer due  
to subsurface mixing induced by wave breaking.

For the simulation performed at \(\Rey = 40000\) with a wave steepness of 
 0.37 which most closely resembles the microbreaking conditions measured by
\citet{zappa2001} at the lowest wind speed the maximum gas transfer velocity is
found to be \(k_{660} \simeq 33.6\) cm/h, which appears significantly higher
than the experimentally observed values.
However, examining the experimental wave profiles reveals that the microbreaking
fronts are relatively short-crested. Moreover, at the lowest wind speed tested,
the active breaking area represented only about 10\% of the total surface. In
contrast, the numerical simulations owing to the use of periodic boundary
conditions in the spanwise direction effectively model the waves as
two-dimensional or, more precisely, long-crested.
In the simulations, the active breaking area can be identified as the portion of
the air-water interface where the gas flux exceeds a specified threshold. By
setting this threshold equal to the flux level observed in non-breaking waves,
we find that, at the time corresponding to the peak gas transfer velocity,
approximately 85\% of the interface contributes to gas exchange.
Accounting for the difference in active breaking area between the simulations
and the experiments, the adjusted numerical estimate of the gas transfer
velocity becomes approximately 3.95 cm/h well within the range of the
experimental measurements.

\section{Conclusions}

In conclusion, to the authors' knowledge, this study is the first to utilize a multiphase flow solver to compute gas transfer velocity in wave-breaking flows. The results underscore the critical role of air entrainment in enhancing gas transfer during breaking events and emphasize the importance of accounting for the total air-water interface area to achieve accurate estimates of gas transfer velocity, a capability uniquely enabled by multiphase solvers.  
The computed gas transfer velocities scale with the energy dissipation rate to the power of \(1/4\), aligning with the theoretical predictions. This scaling is particularly significant, as the dissipation rate, at least for the potential energy component, can be inferred from free-surface measurements. However, estimating the air-water interface area remains more challenging. A promising approach involves combining the volume flux of air entrainment with bubble size distributions, as suggested in prior studies~\citep{deike2018grl, deane2002}.  
At this stage, it is important to acknowledge the key limitations of this study. First, extending the observed trends to the higher Schmidt and Reynolds numbers characteristic of oceanic conditions requires further investigation.
Additionally, this study does not account for wind effects, which can significantly influence gas transfer through wind stress and turbulence. As shown in related work~\citep{lu2024prf}, incorporating wind introduces additional complexities. Future research will focus on integrating these effects to further refine the understanding of gas transfer dynamics.

%%%%%%%%%%%%%%%%%%%%%%%%%%%%%%%%%%%%%%%%%%%%%%%
%
% DATA SECTION and ACKNOWLEDGMENTS
%
%%%%%%%%%%%%%%%%%%%%%%%%%%%%%%%%%%%%%%%%%%%%%%%

\section*{Open Research Section}
The data presented in this work were obtained using an in-house multiphase solver. 
Details on the solver can be found in \cite{digiorgio2022jfm,digiorgio2024jcp}. 
The data are publicly available in the Zenodo repository~\citep{di_giorgio_2025_zenodo}.

\acknowledgments
We acknowledge that the results reported in this paper have been achieved using
the EuroHPC Research Infrastructure resource LEONARDO based at CINECA,
Casalecchio di Reno, Italy.

\bibliographystyle{plainnat}
\bibliography{reference}

\end{document}